\documentclass{PoS}
\usepackage{graphicx}

\title{Multi-Wavelength Properties of Radio-loud Narrow-Line Seyfert 1 Galaxies}

\ShortTitle{Radio-loud NLS1 galaxies}

\author{\speaker{S. Komossa}  \\
        Max-Planck-Institut f{\"u}r Radioastronomie, Auf dem H{\"u}gel 69, 53111 Bonn, Germany\\
        E-mail: \email{astrokomossa@gmx.de}}

\abstract{Radio-loud Narrow-line Seyfert 1 (NLS1) galaxies are unique probes 
of the formation of powerful radio jets at extreme (near-Eddington) accretion rates and low black hole
masses, in a regime very different from classical blazars and, therefore, not probed before. Further, strong thermal and non-thermal
components simultaneously present in their spectral energy distributions (SEDs), 
shed new light on the jet-disk connection at much shorter timescales, given their low SMBH masses.
While NLS1 galaxies have been studied thoroughly at optical and X-ray energies
for decades, the populations of radio-loud, and $\gamma$-ray emitting NLS1 galaxies 
have only emerged more recently. This contribution provides a short review of the multi-wavelength
properties of this intriguing class of Active Galactic Nuclei (AGN), including X-ray variability and emission mechanisms, $\gamma$-ray discoveries, SEDs, star formation activity, host galaxy morphologies, different methods of black hole mass estimation, (unexpected) FeII properties, and high-velocity large-scale outflows possibly driven by the radio jet and with a surprisingly pronounced ionization stratification.   
}

\FullConference{Revisiting narrow-line Seyfert 1 galaxies and their place in the Universe - NLS1-2018\\
		9-13 April 2018 \\
		Padova Botanical Garden, Italy}

\begin{document}

\section{Introduction}

Narrow-line Seyfert 1 (NLS1) galaxies are a subclass of Active Galactic Nuclei (AGN) with extreme multi-wavelength
properties. The criteria commonly used to define narrow-line type 1 AGN are small widths of their broad Balmer lines of
FWHM(H$\beta$) $<$ 2000 km/s, faint emission of [OIII]/H$\beta < 3$,
and the presence of strong FeII emission complexes 
(\cite{osterbrock85}, \cite{goodrich89},
\cite{veron-cetty01}). The FeII emission anti-correlates in strength with the [OIII] emission, and with the width of the broad Balmer lines.
Their properties place NLS1 galaxies at one extreme end of AGN correlation space, and NLS1 galaxies therefore provide us with important insights in the drivers and physics of AGN.  

While their characteristic optical properties were recognized early in individual NLS1 galaxies (e.g., \cite{sargent68}, \cite{koski78}), they were introduced as a subclass of AGN by Osterbrock \& Pogge (\cite{osterbrock85}). The term ``narrow line Seyfert 1'' was coined by Gaskell (\cite{gaskell84}). 
Larger numbers of NLS1 galaxies were identified in spectroscopic follow-ups of bright ROSAT X-ray sources (\cite{grupe99}, \cite{xu03}). More recently, the outstanding spectroscopic data base of the Sloan Digital Sky Survey (SDSS; \cite{york00}) enabled the selection of large NLS1 samples (\cite{williams02}, \cite{zhou06}, \cite{rakshit17}).  

The small width of their broad Balmer lines, FWHM(H$\beta$) $<$ 2000 km/s, has become a convenient selection criterion of NLS1 galaxies. While there is no sharp change at 2000 km/s{\footnote{Some authors prefer the use of 4000 km/s, instead, and then distinguish between ``Population A and B'' AGN (e.g., \cite{sulentic00}, \cite{marziani18}). It is also well known that, ideally, the FWHM-cut should increase with absolute magnitude, since the luminous quasars generally have broader emission-line widths.}}, AGN properties, statistically, do change as a function of broad-line width with NLS1 galaxies lying at one extreme end of AGN parameter space. Principal component analysis and other correlation analyses have revealed that, on average as a class, NLS1 galaxies show stronger FeII emission, steeper soft X-ray spectra, stronger X-ray variability, lower narrow-line region (NLR) density, and higher-velocity outflow components in their NLRs than broad-line Seyfert 1 (BLS1) galaxies (e.g., \cite{boroson92}, \cite{boller96}, \cite{leighly99}, \cite{xu07}, \cite{komossa08b}, respectively).  

\begin{figure*}[b]
\centering
\includegraphics[width=3.9in]{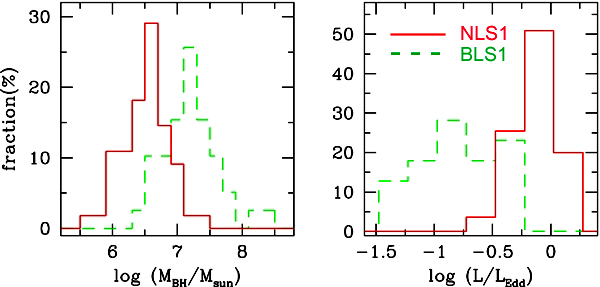}
\caption{Distribution of black hole masses and Eddington ratios of a sample of (radio-quiet) NLS1 and BLS1 galaxies (\cite{xu12}).}
\label{masses}       
\end{figure*}

Some of the extreme properties of NLS1 galaxies can be understood in terms of near-Eddington accretion onto relatively low-mass black holes (e.g., \cite{boroson02}, \cite{grupe04}, \cite{grupe10},
\cite{xu12}; review by \cite{komossa08}). The low black hole masses explain the small widths of the broad-line region (BLR)
emission lines. Along with their multi-wavelength emission (and evidence for strong EUV bumps in their SEDs), this then implies accretion at or near the Eddington limit
(Fig. \ref{masses}). NLS1 galaxies are, therefore, rapidly growing their supermassive black holes (SMBHs), and may represent young or re-juvenated AGN (\cite{mathur00}).
In the Eigenvector (EV) analysis of \cite{boroson02},
the Eddington ratio $L_{\rm bol}$/$L_{\rm Edd}$ drives the first eigenvector EV1, while accretion rate drives EV2. 

While NLS1 galaxies are on average more radio-quiet
than broad-line AGN, 
a fraction of NLS1s are beamed and radio-loud,  
are highly variable at radio frequencies, and 
are detected at $\gamma$-rays.  
This contribution provides a review of the multi-wavelength properties of this remarkable new class of radio-loud, jet-emitting AGN. 

\section{Identification of radio-loud NLS1 galaxies}

``{\it{Radio-loud (RL) sources are found to be an AGN population with fundamental geometrical and kinematic difference from the radio-quiet (RQ) majority. NLSy1 sources are found to be an extremum of the RQ
population.}}'' --- Sulentic et al. (2000) \cite{sulentic00}

\vskip0.3cm

Radio-loudness of an AGN is commonly measured based on the radio index $R$; the ratio of the 6 cm radio flux over the optical flux at 4400\AA, following \cite{kellermann89}. A value of $R=10$
is then frequently used to mark the formal border between radio-loud and radio-quiet. 
$R$ was initially calculated based on the assumption that 
spectral slopes in the radio and optical are similar, with $\alpha =-0.5$. With these assumptions, the radio index $R_{1.4}$ at 1.4 GHz (where the FIRST and NVSS
surveys measured) is then given by $R_{1.4}=1.9R$. The radio-loudness ratio $R$ is only well defined in the absence of optical extinction.
Alternatively, the radio power, $P_{\rm 5\;GHz} > 10^{24}$ W/Hz (\cite{joly91}), has been used as an order-of-magnitude
discriminator between radio-loud and radio-quiet, independent of the optical emission. 

While NLS1 galaxies have been intensely studied in the optical and X-ray domain, little was known about their radio properties prior to $\sim$2005. 
However, a few individual sources were noticed to be radio emitters and radio-loud. PKS0558--504 (\cite{remillard86}),
RXJ0134--4258 (\cite{grupe00}), SDSSJ094857.3 +002225
(\cite{zhou03}), and SDSSJ172206.03+565451.6 (\cite{komossa06b}) all exceed radio indices $R>10$ and show typical optical NLS1 spectra, while PKS2004--447 is very radio-loud, but with an NLS1-untypical optical spectrum (\cite{oshlack01}). Ulvestad et al. (\cite{ulvestad95}) observed a 
mini-sample of seven (radio-quiet) NLS1 galaxies with the Very Large Array (VLA) and concluded that their radio power is
modest, and their spatial extent less than a few hundred parsecs. 

A systematic search for radio-emitting and radio-loud NLS1 galaxies was carried out by Komossa et al. (\cite{komossa06a}). Were these sources truly rare, or had they just been overlooked ? 
Identifying such systems allows us to address several key questions related to our understanding of the physics of radio-emitting 
AGN in general, and NLS1 models in particular, including 

\begin{itemize}
\item { {\bf{(1) Jet-disc coupling:}} What role does the accretion disc play in the launching of jets ?} The SEDs of the blazar population of NLS1 galaxies often show {\em both} components, strong thermal emission from the accretion disc, and non-thermal emission from the jet. While this is also seen in flat-spectrum radio quasars (FSRQs) (\cite{marscher02}), variability timescales in NLS1 galaxies are much shorter, and therefore, they represent ideal laboratories for studying the jet-disc connection.   

\item { {\bf{(2) Radio-loud radio-quiet dichotomy and driver(s) of radio-loudness:}} What makes an object radio-loud ? Why is about 15\% of the AGN population radio-loud, while the rest is radio-quiet or radio-silent ? What is the role of black hole spin and large-scale environment (host galaxy)? NLS1 galaxies span a very different parameter space than classical blazars, allowing us to re-address this key unsolved puzzle in extragalactic astrophysics. }

\item{ {\bf{(3) Orientation models for NLS1 galaxies:}} Are NLS1 galaxies preferentially viewed face on?
(If so, we might expect the fraction of radio-loud sources to be high). Can this explain some of their intriguing multi-wavelength properties?  Does this affect their SMBH mass determinations?}

\end{itemize}

\begin{figure*}[t]
\centering
\includegraphics[width=3.4in]{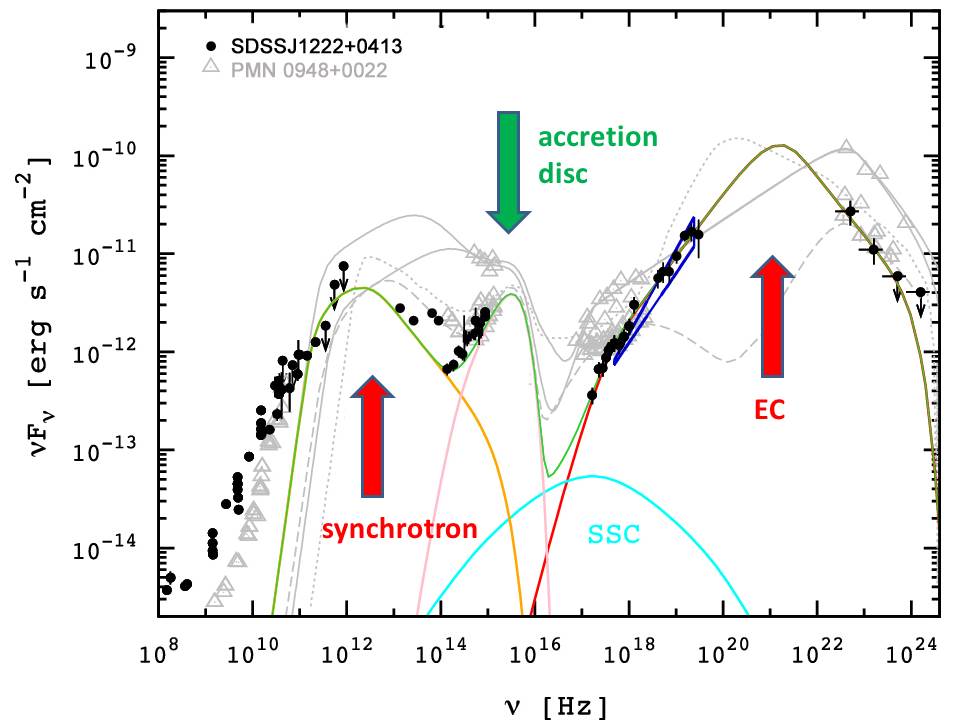}
\caption{Example of a typical SED of a radio-loud NLS1 galaxy (taken from \cite{yao15b}). In addition to the
characteristic double-hump structure of blazars, a strong accretion-disc contribution in the optical--EUV--soft-X-rays is
often present. Along with their small SMBH masses, i.e., more rapid timescales of variability, this makes radio-loud NLS1
galaxies excellently suited for studying the disc-jet symbiosis. The black symbols represent the SED of SDSSJ1222+0413 (\cite{yao15b}), while the grey symbols are measurements and models of PMN0948+0022 at different epochs, taken from \cite{dammando15b}, \cite{sun15} and \cite{foschini12}.}
\label{sed-general}       
\end{figure*}

\begin{figure*}[t]
\centering
\includegraphics[width=3.5in]{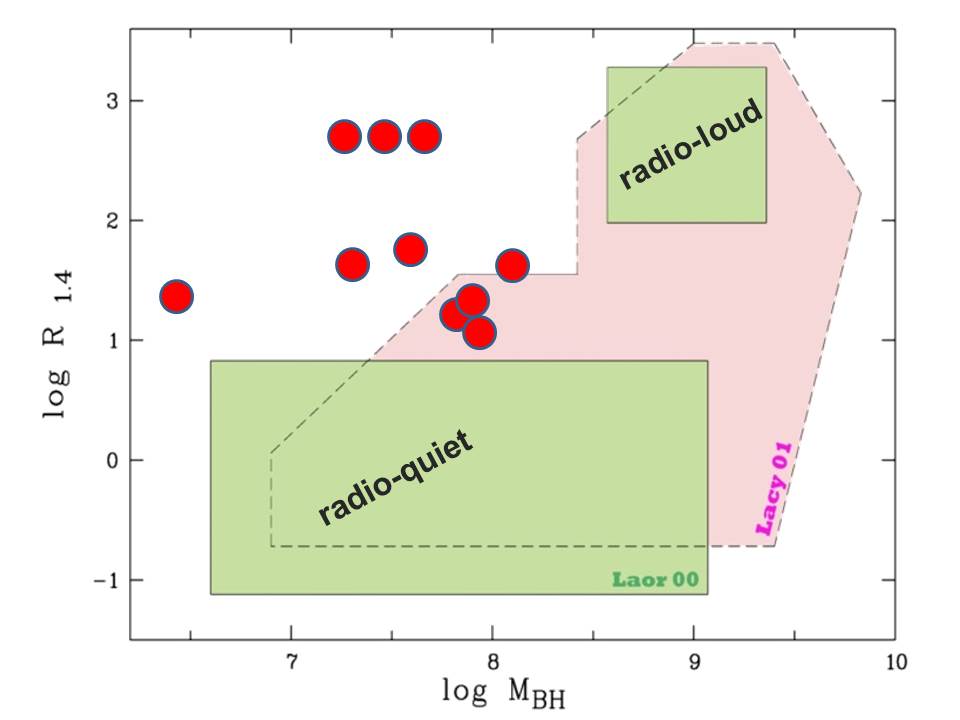}
\caption{Distribution of radio-loud NLS1 galaxies (red circles) in the SMBH mass -- radio-loudness diagram (\cite{komossa06a}; see also Fig. 16 of \cite{yuan08} and Fig. 6 of \cite{jaervelae15}). Their low SMBH masses and high radio-loudnesses make NLS1s stand out.
They are therefore unique probes of jet activity in previously barely explored parameter space. The two
shaded areas are from BLS1 samples of \cite{laor00} and \cite{lacy01}. }
\label{radio-loud-results}       
\end{figure*}

The sample of Komossa et al. (\cite{komossa06a}) was composed of 
{\em quasars} (brighter than $M=-23$ mag), drawn from the AGN catalogue of Veron-Cetty \& Veron (\cite{veron-cetty03}), cross-correlated with
various available Northern and Southern radio catalogues. Even though we use the term {\em quasars}, here, we generally do not 
distinguish between narrow-line type 1 quasars, and narrow-line Seyferts 1s, but call all sources ``NLS1s'' for the sake of brevity. The main findings can be summarized as follows: 

\begin{itemize}

\item {Only 7\% of the NLS1 galaxies are radio-loud (R>10).}

\item {Only 2.5\% are very radio-loud (R>100), while 14\% of the BLAGN of the same study are very radio-loud.} 

\item {SMBH masses are {\em much lower} than classical radio-loud AGN (even though at the upper end of the NLS1 distribution). These sources are therefore in a previously {\em rarely covered} parameter space (e.g., \cite{laor00}) of low SMBH masses and high radio-loudness (Fig. \ref{radio-loud-results}).}

\item {All radio-loud NLS1 galaxies turned out to be {\em strong FeII emitters}, even though FeII strength and radio-loudness were previously found to be at opposite ends of AGN parameter space (e.g., \cite{sulentic00}; see our Sect. 7.3 and Fig. \ref{FeII}).} 

\item {Eddington ratios $L/L_{\rm Edd}$ are high.}

\item {~70\% of the sources of the sample are compact radio-emitters (unresolved with FIRST) with steep spectra, therefore CSS-like (i.e, share similarities with compact steep-spectrum sources).}

\item {~30\% are blazar-like, with flat/inverted radio spectra and/or strong radio variability.}

\end{itemize}

The low fraction of radio-loud narrow-line type 1 AGN has been confirmed in other samples. \cite{chen18} reported a fraction of 6\% radio-loud NLS1 galaxies in a Southern sample, selected from 6dFGS, and \cite{cracco16} found 4\% radio-louds in nearby NLS1 galaxies from SDSS. \cite{zhou06} mentioned 6\% radio-detections in SDSS-selected NLS1 galaxies, compared to 5\% detections by \cite{rakshit17}. In that latter sample, only 3.7\% are radio-loud (\cite{singh18}).    

\cite{komossa06a} focussed on {\em quasars}, because radio-loudness can be much more reliably determined (e.g., pointlike hosts, fiber centered well on core; and with little or no extinction). 
In order to exclude uncertainties in radio-loudness determination near the border of $R=10$ when including low-luminosity, nearby {\em Seyfert} galaxies, \cite{yuan08} in their search for radio-loud NLS1s, therefore, only looked at the radio-loudest population ($R>100$), selected among NLS1 galaxies from SDSS. In that sample, the fraction of blazar-like NLS1 galaxies is even higher, and these sources showed flat or inverted radio spectra, radio variability, enhanced optical continuum emission and blazar-like SEDs, hinting at a sub-population of NLS1 galaxies with relativistic jets. A third sample of radio-emitting NLS1 galaxies was identified by \cite{whalen06} in optical follow-ups of radio sources, who concluded that optical properties of radio-loud and radio-quiet NLS1 galaxies are overall similar.    

Many follow-up studies in the radio regime of the radio-loud population, either sub-samples, or individual galaxies, have been carried out in recent years (since
there is a separate review on radio properties by \cite{lister18}, we only give a very condensed summary here). 
High-resolution radio imaging has revealed generally compact morphologies, compact cores, 
one-sided core-jet structures on parsec scales, several sources with two-sided jets, a few with faint, widely extended
emission on kpc scales, and core brightness temperatures typically $<10^{11}$ K, less than the classical blazar
population with $\sim 10^{11-13}$ K 
(e.g., \cite{doi06}, \cite{doi07}, \cite{doi11}, \cite{doi12}, \cite{doi18}, \cite{giroletti11}, \cite{dammando12}, \cite{dammando13}, \cite{wajima14}, \cite{richards15}, \cite{gu15}, \cite{schulz16}, \cite{fuhrmann16}, \cite{lister16}, \cite{berton18a}, \cite{singh18}).  

Several of the $\gamma$-ray emitting NLS1 galaxies exhibit superluminal motion [e.g., SBS0846+513: 8.2c (\cite{dammando12}); 1H0323+342: up to 7c (\cite{fuhrmann16}); PKS 1502+036: 1.1c, \cite{lister16})].  
Combining imaging and monitoring radio data, 1H0323+342 was used for the first direct viewing angle determination toward a NLS1 galaxy, $\Theta < 4^{\rm o}-13^{\rm o}$ (\cite{fuhrmann16}).  
\cite{doi18} presented evidence for a recollimation shock in the jet of 1H0323+342 and suggested that the quasi-stationary jet feature is one of the possible $\gamma$-ray emitting sites. 

As expected, the blazar population among the radio-loud NLS1 galaxies is highly variable in the radio regime, with rapid and repeat flaring, spectral variability more pronounced at higher frequencies, and moderate variability brightness temperatures (e.g., \cite{abdo09c}, \cite{foschini12}, \cite{dammando12}, \cite{paliya13}, \cite{angelakis15}, \cite{schulz16}, \cite{l17}).

\section{X-ray properties: disc and jet contributions}

Radio-quiet NLS1 galaxies, as a class, are known for their steep X-ray spectra in the soft X-ray band (0.1-2 keV), and many of the bright soft X-ray AGN identified with ROSAT turned out to be NLS1 galaxies (\cite{grupe99}). 
NLS1 galaxies are highly variable in the X-ray regime, and several of them underwent high-amplitude outbursts or reached deep low states
(e.g., WPVS007: \cite{grupe08a}, 1H0707--495: \cite{fabian12}, Mrk\,335: \cite{grupe08b}, IRAS13224--3809: \cite{parker17}).
Their X-ray spectra show strong signs of reprocessing, and they are therefore valuable tools for probing the physics of matter in the SMBH environment of AGN (see the review by \cite{gallo18} in these proceedings). 
Mrk\,335, even though radio quiet, shows X-ray events -- characteristic changes in the geometry of the corona inferred from relativistically blurred reflection off the accretion disc -- which are well explained by an aborted jet-launching event (\cite{wilkins15}). 

\begin{figure*}[t]
\centering
\includegraphics[width=5.0in]{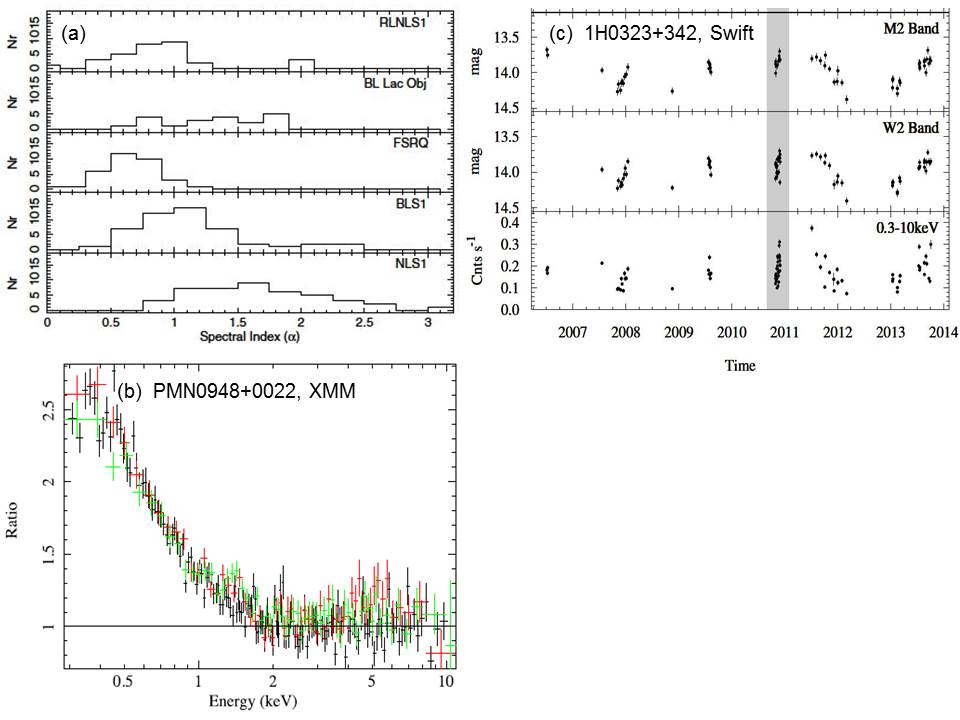}
\caption{(a) Average X-ray spectral indices $\alpha_{\rm X}$ of radio-loud NLS1 galaxies, in comparison with other
source types (taken from \cite{foschini15}). (b) XMM-Newton spectrum of PMN0948+0022, displayed as ratio spectrum over a hard powerlaw (\cite{dammando14}). The soft X-ray emission is well modelled by Comptonization of disc photons or blurred reflection, while jet emission dominates above $\sim$2 keV. (c) High-amplitude X-ray variability of 1H0323+342 (\cite{yao15a}; see also \cite{paliya14}) detected by the {\sl{Neil Gehrels Swift}} observatory.}
\label{Xrays}       
\end{figure*}

\subsection{Sample properties}
Regarding radio-loud NLS1 galaxies, the X-ray band (0.1-10 keV) is an interesting regime, where both accretion discs and jets potentially make strong
contributions (thermal or reprocessed soft X-rays from the disc and harder X-rays from the corona above the disc; SSC and EC
emission from the jet). In the soft X-ray band (0.1-2.4 keV), the spectra of radio-loud NLS1 galaxies are flatter than their radio-quiet counterparts, with average photon indices  $<\Gamma_{\rm x}>$ = --2.4 (in both samples of \cite{komossa06a} and \cite{yuan08}).
Foschini et al. (\cite{foschini15}), in their systematic study of the X-ray properties of radio-loud NLS1 galaxies, found $<\Gamma_{\rm x}>$ = --2.0 in the  (0.3-10)\,keV band, and $<\Gamma_{\rm x}>$ = --2.7  and --1.6, in comparison samples of radio-quiet NLS1 galaxies and FSRQs, respectively (Fig. \ref{Xrays}a). 

These results suggest that thermal, disc-related components are still present in radio-loud NLS1 galaxies but are less pronounced than in the radio-quiet population, and are more pronounced than in the FSRQ population. 

\subsection{Spectroscopy and variability}

Deep X-ray spectroscopy with the X-ray observatories XMM-Newton, Suzaku, and NuSTAR of selected sources
has revealed that most radio-loud NLS1 galaxies studied thus far show significant soft X-ray emission 
components, that are most plausibly related to the accretion disc (e.g., \cite{gallo06}, 
\cite{derosa08}, 
\cite{gliozzi13},
\cite{dammando14}, 
\cite{paliya14}, 
\cite{orienti15}, 
\cite{yao15a}, 
\cite{ghosh16}, 
\cite{landt17}, 
\cite{kynoch18}, 
\cite{ghosh18}, 
\cite{larsson18}; 
Fig. \ref{Xrays}b), but apparently absent in a few cases, at least at some epochs (PKS2004--447: \cite{orienti15}, \cite{kreikenbohm16}; SDSSJ2118--0732: \cite{yang18}). 

The mildy radio-loud NLS1 galaxy PKS0558--504 has been extensively monitored in X-rays and is highly variable (e.g., \cite{gliozzi07}). It seems to be jet-dominated only in the radio regime. Its X-ray spectrum shows a soft excess (\cite{ghosh16}) which likely
provides the seed photons for the hard coronal X-ray emission (\cite{gliozzi13}). RXJ0134--4258, another NLS1 galaxy that was identified as radio-loud early (\cite{grupe00}), underwent
dramatic X-ray spectral variability (\cite{komossa00}), changing from super-steep ($\Gamma_{\rm x}=-4.4$) to unusually flat ($\Gamma_{\rm x}=-2.2$), well modelled by variable ionized
absorbing material of high column density. 

Rapid X-ray variability, sometimes down to timescales of hours, 
is common in radio-loud NLS1 galaxies. 
1H0323+342, monitored by the {\sl{Neil Gehrels Swift}} observatory (Swift, hereafter) for $>$10 years by now (Fig. \ref{Xrays}c), is highly variable in all wavebands, and exhibits repeat rapid flaring in X-rays. 
UV and X-ray variability is correlated in 1H0323+342, and consistent with a disc origin (\cite{yao15a}). 
Its X-ray spectrum is the only one among the radio-loud NLS1s which is known to show evidence for a weak Fe emission line (\cite{paliya14}, \cite{kynoch18}, \cite{ghosh18}). 

\begin{figure*}[t]
\centering
\includegraphics[width=5.1in]{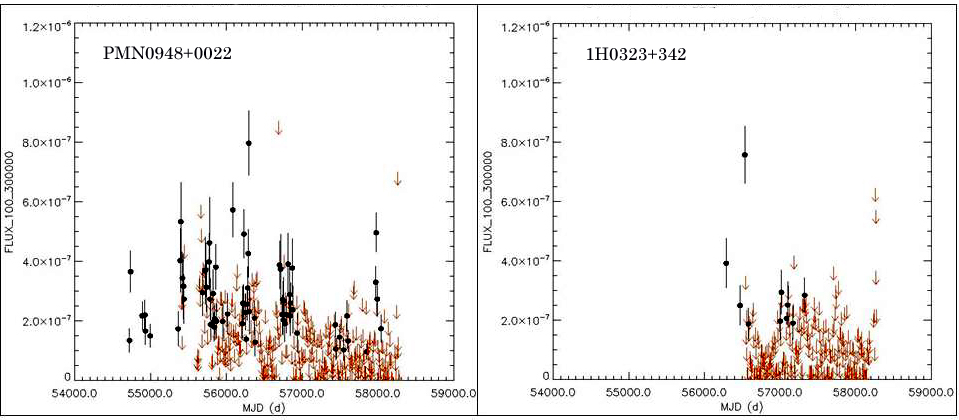}
\caption{Fermi $\gamma$-ray lightcurves of PMN0948+0022 and 1H0323+342 (0.1-300 GeV; weekly time bins), as provided at, and directly retrieved from, the Fermi webpage at https://fermi.gsfc.nasa.gov/ssc/data/access/lat. Both sources have been relatively inactive during the last year.}       
\end{figure*}

\section{$\gamma$-ray detection of radio-loud NLS1 galaxies}

With the Large Area Telescope (LAT) on board the $\gamma$-ray satellite Fermi, $\gamma$-ray emission from several NLS1 galaxies was discovered 
(\cite{abdo09a}, \cite{abdo09b}), and more identifications followed (\cite{foschini11}, \cite{dammando12}, \cite{dammando15a}, \cite{yao15b}, 
\cite{yang18}, \cite{paliya18}; Tab. 1 and Fig. 5). Sources show: 

\begin{itemize}
\item {Rapid and repeated flaring detected in several of the $\gamma$-ray lightcurves on timescales as short as $\Delta t \sim$3--30 d (e.g., PMN0948+0022, SBS0846+513, 1H0323+342, PKS1502+036, and SDSSJJ2118--0732; Fig. 5).}

\item {High isotropic $\gamma$-ray luminosities up to $L_{\rm peak} > 10^{48}$ erg/s (PMN0948+0022).} 

\item {These observations have established NLS1 galaxies as a new group of $\gamma$-ray emitting AGN, and  
re-confirmed their blazar-like nature
and the presence of powerful (relativistic) jets. It is currently under investigation, whether they represent low-mass analogues of FSRQs, or whether their jets are intrinsically different. } 

\end{itemize} 


\begin{table*}[t]
\small
\begin{center}
\begin{tabular}{lllccl}
\hline
galaxy name & coordinates & & redshift & radio index $R$ & reference \\
            & RA  & DEC     &          &     & \\
\hline
\hline
PMN\,0948+0022 & 09:48:57.32 & +00:22:25.5 & 0.585 & 350 & \cite{abdo09a} \\
1H0323+342   & 03:24:41.19 & +34:10:45.9 & 0.063 & 50 & \cite{abdo09b} \\
PKS\,1502+036  & 15:05:06.48 & +03:26:30.8 & 0.409 & 1550 &  \cite{abdo09b}\\
PKS\,2004--447$^1$ & 20:07:55.18 & --44:34:44.3 & 0.24 & 1710 & \cite{abdo09b} \\
SBS\,0846+513  & 08:49:57.97 & +51:08:29.0 & 0.584 & 1450 & \cite{foschini11}, \cite{dammando12} \\ 
FBQSJ1644+2619 & 16:44:42.53 & +26:19:13.2 & 0.145 & 450 & \cite{dammando15a} \\
SDSSJ1222+0413$^*$ & 12:22:22.55 & +04:13:15.7 & 0.966 & 3230 & \cite{yao15b} \\
SDSSJ2118--0732$^1$ & 21:18:52.96 & --07:32:27.5 & 0.26 & 920 & \cite{yang18}, \cite{paliya18} \\
SDSSJ0932+5306 & 09:32:41.1 & +53:06:33.3 & 0.60 &  & \cite{paliya18} \\
GB6\,J0937+5008$^1$$^*$ & 09:37:12.33 & +50:08:52.1 & 0.28 & &  \cite{paliya18} \\
SDSSJ0958+3224 & 09:58:20.9 & +32:24:01.6 & 0.53 &  & \cite{paliya18} \\
SDSSJ1421+3855$^2$ & 14:21:06.0 & +38:55:22.5 & 0.49 &  & \cite{paliya18} \\
TXS\,1518+423$^*$   & 15:20:39.61 & +42:11:08.9 & 0.48 &  & \cite{paliya18} \\
PMN\,J2118+0013$^*$ & 21:18:17.40 & +00:13:16.8 & 0.46 &  & \cite{paliya18} \\
SDSSJ1641+3454$^{3}$ & 16:41:00.10 & +34:54:52.7 & 0.164 & 13 & \cite{l18} \\
\hline
\end{tabular}
\caption{$\gamma$-ray emitting NLS1 galaxies, detected at high significance and published in main journals (see \cite{foschini11}, \cite{liao15},
\cite{dammando17}, \cite{miller17}, \cite{berton17} and \cite{l18} for further candidates). Coordinates (RA, DEC) in J2000. 
\newline
\noindent $^1$Little or no optical FeII emission. $^2$FWHM(H$\beta$) not well measured, therefore NLS1 classification remains uncertain (\cite{paliya18}). $^3$Very red optical spectrum. $^*$Already known as blazar and/or $\gamma$-ray emitter (\cite{ackermann15}), but later optically re-classified as NLS1 galaxy.  
}
\label{gammas}
\end{center}
\end{table*}
\normalsize

Because of the flaring, some sources are only detected occasionally or just once with Fermi, making counterpart identification and confirmation more challenging, given
the large location uncertainties of Fermi.  
Nevertheless, we expect more sources to appear, as Fermi continues its measurements. Evidence for faint and/or sporadic
$\gamma$-ray emission was reported for several more radio-loud NLS1 galaxies (\cite{foschini11}, \cite{liao15},
\cite{dammando17}, \cite{miller17}, \cite{l18}). \cite{berton17} recently re-classified the optical spectrum of the quasar 3C286 as a narrow-line type 1 AGN. 3C286, as they point out, is
also identified as a $\gamma$-ray emitter in the 3rd Fermi catalogue (3FGL\,J1330.5+3023; \cite{ackermann15}).   

Among these candidates, several NLS1s have relatively steep radio spectra.
PKS2004--447 (\cite{oshlack01}, \cite{gallo06}, \cite{schulz16}),  
RXJ2314.9+2243 (\cite{komossa15}), B3\,1441+476 (\cite{liao15}),
and 3C286 (\cite{berton17}) are steep-spectrum sources  -- at least most of the time{\footnote{note, that strong spectral variability in blazars sometimes leads to temporary type-changes from flat-spectrum to 
steep-spectrum}}. The unusual multi-wavelength properties of RXJ2314.9 +2243 stand out and were discussed in more detail;   
including its UV--IR SED, high-velocity outflow traced by [OIII]$\lambda$5007 ($\Delta v_{\rm [OIII], wing} = 1260$ km/s), and its constant radio emission at multiple epochs and multiple frequencies observed with the Effelsberg telescope  (\cite{komossa15}), despite its candidate $\gamma$-ray detection. 

Given a population of ($\gamma$-ray emitting) NLS1 galaxies with jets pointing toward us, the question is raised about the parent population of these systems, seen at high inclination angle. Generally, the CSS-like NLS1 galaxies which have been identified in the early NLS1 samples (\cite{komossa06a}, \cite{yuan08}, \cite{gu15}), were discussed in a sequence of papers by Berton and collaborators (e.g., \cite{berton15}, \cite{berton16a}, \cite{berton17}; see also \cite{caccianiga14}), as part of that parent population -- likely including the four candidate $\gamma$-ray NLS1 galaxies mentioned in the previous paragraph.   

\section{SED modelling}
Blazars show a double-humped SED dominated by non-thermal emission from the jet. At low energies, synchrotron emission is the main radiation mechanism, while at higher energies synchrotron self-compton (SSC) emission from jet photons, or external Comptonization (EC) of seed photons from the accretion disc, torus or BLR, dominates (see the reviews by \cite{maraschi08} and \cite{boettcher12}).  

The $\gamma$-ray emitting NLS1 galaxies were observed in a number of multi-wavelength campaigns and therefore generally have the best-covered SEDs among the radio-loud population. 
Their SEDs show the characteristic double-hump structure of blazars with a high Compton dominance 
(Fig. 2). In many cases, an accretion-disc component is additionally present. 
The SEDs are generally well modelled with one-zone leptonic jet models with external Comptonization (e.g., \cite{abdo09a}, \cite{abdo09b}, \cite{abdo09c}, 
\cite{foschini11}, \cite{dammando12}, \cite{dammando13}, \cite{dammando15b}, \cite{dammando16}, 
\cite{foschini11b}, \cite{foschini12}, \cite{foschini15}, \cite{paliya13}, \cite{paliya14}, \cite{paliya18}, \cite{zhang13}, \cite{maune14}, \cite{sun15},  
\cite{yao15a}, \cite{yao15b}, \cite{orienti15}, \cite{komossa15}, \cite{yang18}). Based on detailed
modelling of the SED of 1H0323+342, \cite{kynoch18} concluded that the site of the $\gamma$-ray formation is located inside the BLR, and that the external Comptonization is due to seed photons from the accretion disc. 

Radio-loud NLS1 galaxies overall have lower jet powers than FSRQs and BL Lacs, but similar values, once scaled by mass (\cite{foschini15}). Other factors which have been suggested to contribute to the lower jet powers of NLS1 galaxies include expansion in a dense medium, repeat jet activity and/or aborted jets, and the young age of the sources.   

\section{Infrared emission: Star formation and jet activity}

Both, AGN jets and/or star formation may contribute to the IR and radio emission of radio-loud NLS1 galaxies. In particular, since radio-quiet NLS1 galaxies often show strong star formation activity (e.g., \cite{sani10}), it is important to explore the role of star formation in radio-loud sources. 
The tight correlation between IR luminosity and radio luminosity of starburst galaxies (\cite{yun01}, their equ. 4) can be used to predict the radio emission expected from starbursts. 
This relation has been used to show that the radio emission of radio-loud NLS1s is up to several orders of magnitude more
luminous than expected from starbursts and therefore AGN-jet-related (\cite{komossa06a}). 
\cite{caccianiga15} used WISE and presented some radio-loud NLS1 galaxies which are likely predominantly 
powered by a starburst. The majority of the sources in their sample are still jet dominated. 

With WISE, rapid and even intraday NIR variability was detected in several cases (\cite{abdo09b}, \cite{jiang12}, \cite{yao15b}, \cite{yang18}). Variability of the two fastest sources,
SBS0846+513 and PMN0948+0022, implies size scales $<10^{-3}$ pc, smaller than the torus and consistent with the base of a jet (\cite{jiang12}). Radio-detected NLS1 galaxies are generally more variable in the (WISE) NIR than non-detected ones (\cite{rakshit17}).  

\section{Inferences from optical imaging and spectroscopy}

\subsection{Host galaxies}

Relatively few large-sample studies of the host galaxies of radio-quiet NLS1 galaxies exist, and even less 
is known about the hosts of radio-loud NLS1 galaxies (many of which are, formally, quasars rather than Seyferts). 
Knowledge of their host galaxies is of great interest, since host morphology and host brightness profile, and merging, may play a
significant role in determining whether a quasar
is radio loud (e.g., \cite{capetti06}, \cite{kellermann16}), and since both, mergers and bars, are known to trigger nuclear activity.  

Analyses of samples of nearby (radio-quiet) NLS1 galaxies concluded that there is no excess of companion galaxies and no strong evidence for ongoing mergers (\cite{xu12}, \cite{krongold01}, respectively).
Pseudo bulges are common (\cite{oran11}, \cite{mathur12}). Pre-selecting very nearby spiral host galaxies ($z<0.07$), NLS1 spirals show a higher fraction of bars and nuclear dust spirals, than BLS1 spiral galaxies (\cite{crenshaw03}, \cite{deo06}, \cite{ohta07}). These observations point to the absence of merger-induced accretion, but favor bar-driven inflows and SMBH fuelling in the nearby (radio-quiet) NLS1 galaxies.

\begin{figure*}[t]
\centering
\includegraphics[width=5.9in]{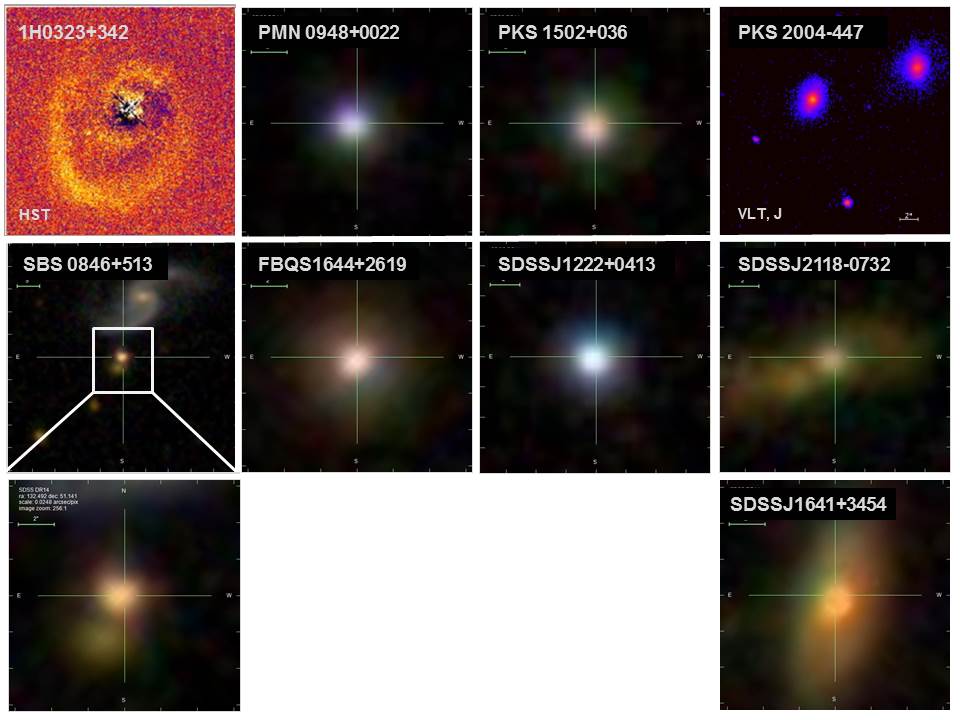}
\caption{Host galaxy images of nine $\gamma$-ray detected NLS1 galaxies, retrieved from the SDSS archive, except for 1H0323+342 (HST image of \cite{zhou07}) and PKS2004--447 (VLT J-band image of \cite{kotilainen16}). The length of each horizontal bar represents a scale of 5 arcseconds in the case of SBS0846+513 (upper panel), and 2 arseconds in all other images. }
\label{hosts}      
\end{figure*}

Very recently, imaging results for several of the $\gamma$-ray NLS1 galaxies have become available.
The host galaxy of the nearest $\gamma$-ray emitting NLS1 galaxy, 1H0323+342 at $z=0.063$, is well resolved with HST and ground-based telescopes
and shows a one-armed spiral or ring-like morphology (\cite{zhou07}, \cite{anton08}, \cite{leon14}; Fig. \ref{hosts}), possibly from a past merger.
SDSSJ2118--0732 is an interacting or merging system (\cite{paliya18},
\cite{yang18}) and the same likely holds for SBS0846+533 (Fig. \ref{hosts}). Imaging of PKS2004--447 (\cite{kotilainen16}) revealed a (pseudo)bulge plus disc and bar, and no evidence
for a recent merger. \cite{dammando17}, using the GTC\,10m telescope, reported evidence for an
elliptical host of FBQSJ1644+2619, while \cite{olguin-iglesias} prefer a barred
lenticular (SB0) host galaxy model of FBQSJ1644+2619, with a pseudo-bulge and a ring, perhaps the aftermath of a minor merger. The host morphology of PKS1502+036 is well fit with a Sersic index $n=3.5$, indicative of an elliptical host galaxy (\cite{dammando18}), and there is some evidence for a galaxy interaction.   
Imaging of the mildly radio-loud IRAS20181--2244 has revealed an interacting or merging pair of galaxies with a pronounced tidal tail; the host galaxy is well described by a disc morphology (\cite{berton18b}). 

We have visually inspected all SDSS images of the remaining radio-loud (non-gamma-emitting, non-reddened) NLS1 galaxies of the samples of \cite{komossa06a} and \cite{yuan08} at $z<0.4$. All appear unresolved in SDSS with very few exceptions: SDSSJ163323.58+471859.0 ($z=0.116$) is widely extended and appears to host a double-nucleus, or is in an interacting system. SDSSJ172206.03+565451.6 ($z=0.425$) shows some asymmetry possibly from interaction/merging (\cite{komossa06b}), and RXJ23149+2243 ($z=0.168$) is extended (\cite{komossa15}). 

The presence of interaction and mergers points to merger-driven fuelling, while bars drive
a strong secular evolution. Both processes, therefore, seem to play a role in powering radio-loud NLS1 galaxies.  Because many of the radio-loud
and $\gamma$-ray detected systems are at higher redshift ($z > 0.4$, and are formally quasars), their
hosts are not resolved with SDSS, and HST or AO-assisted ground-based observations are needed to
make further progress.  

\subsection{Black hole masses}

When estimating virial SMBH masses of NLS1 galaxies from single-epoch spectra in the same way successfully done for BLS1 galaxies (review by 
\cite{peterson14}), then masses of radio-quiet and radio-loud NLS1 galaxies, as a class, turn out to be significantly lower than their BLS1/blazar
counterparts (Sects. 1 \& 2, and Figs. \ref{masses} \& \ref{radio-loud-results}). 
The SMBH mass is given by 
\begin{equation}
M_{\rm BH} = f_{\rm BLR}~~ \frac{R_{\rm BLR} ~~\Delta v^2} {G}, 
\end{equation}
where $R_{\rm BLR}$ is the radius of the BLR, and the velocity dispersion $\Delta v$ is determined from the width of a broad emission line; often H$\beta$. 
FWHM(H$\beta$) potentially depending on projection effects, as parametrized by the inclination-dependent factor $f_{\rm BLR}$ (equ. 7.1), 
projection-independent mass estimates like polarimetry (\cite{savic18}, \cite{popovic18}, \cite{baldi16}) and X-ray variability (\cite{mchardy06}, \cite{nikolajuk09}) are important resources, and
have been employed, too. The following methods of BH mass estimation have been applied to radio-loud NLS1 galaxies: 

\begin{itemize}
\item {Virial estimates from single-epoch spectra (with or without projection effects), and including using the H$\beta$ line luminosity rather than directly the optical continuum luminosity, since the latter may have a jet contribution.}
\item {Reverberation mapping (only one NLS1, 1H0323+342).} 
\item {Spectropolarimetry (only one candidate NLS1, PKS 2004--447).}
\item {X-ray variability: excess variance, or PSD break    
      frequency (only one NLS1, 1H0323+342).}
\item {SED modelling.}
\item {Host galaxy -- BH scaling relations.} 
\item {Statistical arguments on the frequency of radio-loudness in NLS1 galaxies.}
\end{itemize}

\begin{table*}[b]
\small
\begin{center}
\begin{tabular}{llrl}
\hline
galaxy  & method & BH mass  & reference \\
        &        &  [M$_{\odot}]$     & \\ 
\hline
\hline
 1H0323+342 & single-epoch H$\beta$ & 10$^{7}$ & ~~\cite{zhou07} \\  
            & P$\alpha$         & 2\,10$^{7}$ & ~~\cite{landt17} \\
            & reverberation mapping & 6\,10$^{6}$ & ~~\cite{wang16}, $f_{\rm BLR}=1$ \\
            &                       & 3\,10$^{7}$ & ~~\cite{wang16}, $f_{\rm BLR}=6$  \\
            & (opt--UV) SED modelling & 10$^{7}$ & ~~\cite{abdo09b} \\
            & host galaxy   & 2--4\,10$^{8}$ & ~~\cite{leon14} \\
            & X-ray excess variance & 10$^{7}$ & ~~\cite{yao15a} \\
            & PSD break frequency & 3--8 10$^{6}$ & ~~\cite{pan18} \\
\hline
PKS\,1502+036 & single-epoch H$\beta$ & 4--6\,10$^{6}$ & ~~\cite{yuan08} \\
             & MgII     &       10$^{7}$ & ~~\cite{komossa18} \\
       & (opt--UV) SED modelling & 2\,10$^{7}$ & ~~\cite{abdo09b}\\
       & (opt--UV) SED modelling & 4.5\,10$^{7}$ & ~~\cite{paliya16}\\
       & (opt--UV) SED modelling & 2\,10$^{8}$ & ~~\cite{calderone13} \\
   & (opt--UV) SED modelling & $<$ few\,10$^{7}$ & ~~\cite{dammando16} \\
               & host galaxy & 6\,10$^{8}$ & ~~\cite{dammando18} \\
\hline
\end{tabular}
\caption{Summary of BH mass estimates of 1H0323+342 and PKS\,1502+036.   
\newline
}
\label{tab-masses}
\end{center}
\end{table*}

The nearby NLS1 galaxy 1H0323+342 was studied closely, 
with mass estimates available from several different techniques including orientation-independent methods.
We collected the results from the literature. We also recall that 1H0323+342 is one out of only two NLS1 galaxies for which the viewing angle, $\Theta < 4^{\rm o}-13^{\rm o}$, was determined directly (\cite{fuhrmann16}). \cite{zhou07} estimated $M_{\rm BH} = 10^{7}$ M$_{\odot}$ based on a single-epoch H$\beta$ spectrum. \cite{landt17} used broad P$\alpha$ and derived 
$2\,10^{7}$ M$_{\odot}$.  Reverberation-mapping with the Lijiang telescope gave $6\,10^{6}$ M$_{\odot}$ ($f_{\rm BLR}=1$) or $3\,10^{7}$ M$_{\odot}$ ($f_{\rm BLR}=6$, \cite{wang16}). SED modelling of the optical-UV disc emission was carried out by \cite{abdo09b}, who found $10^{7}$ M$_{\odot}$. 
\cite{leon14} reported a high $10^{8.2-8.6}$ M$_{\odot}$ based on different host galaxy image fits after PSF correction, while the X-ray normalized excess variance (\cite{yao15a}) and the break frequency of the X-ray PSD (\cite{pan18}) confirmed a low BH mass of $10^{7}$ M$_{\odot}$ or less.
The different BH mass estimates for 1H0323+342 and also for PKS\,1502+036 are summarized in Tab. {\ref{tab-masses}}. With a few exceptions, most of these point to low SMBH masses. 

Finally, statistical arguments disfavor (one or several orders of magnitude) higher SMBH masses of radio-loud NLS1 galaxies as a class (\cite{komossa06a}): if NLS1 galaxies and BLS1 galaxies had equal masses intrinsically, but the NLS1 population was viewed more pole-on (and so had narrower Balmer lines for that reason), then we would expect {\em more} radio-loud sources among NLS1 galaxies because the effects of beaming would be maximized. This is 
opposite to what is observed. This statistical argument, therefore, argues against intrinsically equal masses in NLS1 and BLS1 galaxies. 

Also, the different large-scale environments of NLS1 and BLS1 galaxies have been used to argue, that the two populations must be intrinsically different rather than just due to orientation effects (\cite{jaervelae17}).  

\begin{figure*}[t]
\centering
\includegraphics[width=5.8in]{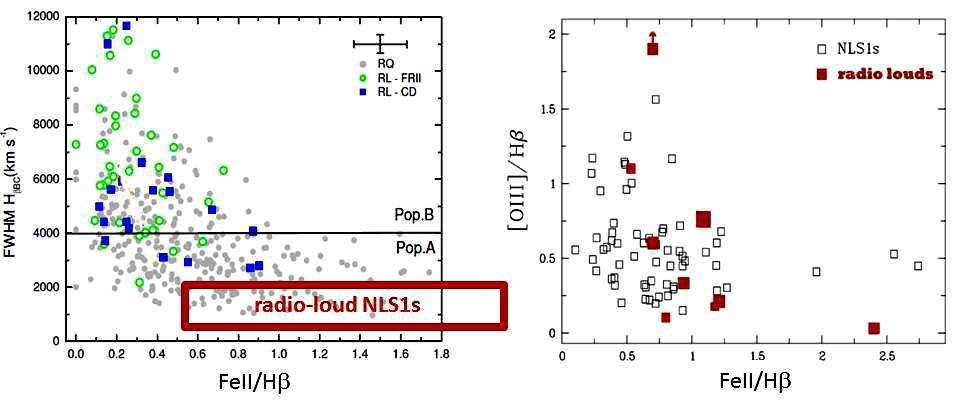}
\caption{Left: FWHM(H$\beta$)--FeII correlation diagram of \cite{sulentic08}. Their radio-loud sources are mostly low-Eddington, high SMBH mass quasars and show {\em only weak} FeII emission (see also Fig. 7 of \cite{boroson02}). Right: [OIII]-FeII diagram of \cite{komossa06a}. Against expectations, the radio-loud NLS1 galaxies all show strong FeII emission and, therefore, populate a previously near-empty ($R$--FeII) parameter space. }
\label{FeII}       
\end{figure*}

\subsection{Emission-line spectroscopy: FeII}

The FeII emission is strong in NLS1 galaxies and is anti-correlated with [OIII] emission and FWHM(H$\beta_{\rm broad}$). This is one of the strongest trends in correlation analyses of quasar spectra (e.g., \cite{gaskell85}, \cite{boroson92}, \cite{sulentic00}, \cite{boroson02}, \cite{grupe04}, \cite{xu07}, \cite{dong11}). 
Due to the complexity of the transitions in the Fe$^+$ ion, the FeII emission complexes consist of a huge number of multiplets. While the bulk of the FeII emission must arise in the BLR, it is not yet clear what powers the line emission (e.g., \cite{baldwin04}). Some NLS1 galaxies exhibit extraordinarily strong FeII emission multiplets which challenge current FeII 
production mechanisms; whether photoionization (radiative heating), or collisional (mechanical heating). 

Interestingly, even though ``FeIIness'' and radio-loudness are at opposite ends of Eigenvector space (e.g., \cite{sulentic00}, \cite{sulentic08}, \cite{boroson02}, \cite{fraix-burnet17}){\footnote{see \cite{sulentic15} for an interesting recent outlier}}, most radio-loud NLS1 galaxies are strong FeII emitters, with FeII/H$\beta$ ratios even larger than a significant fraction of the radio-quiet NLS1 population (\cite{komossa06a}, \cite{yuan08}; Fig. \ref{FeII}). This holds for both radio-loud NLS1 types, the blazar-like sources, and the CSS-like sources. 

\begin{figure*}[b]
\centering
\includegraphics[width=5.7in]{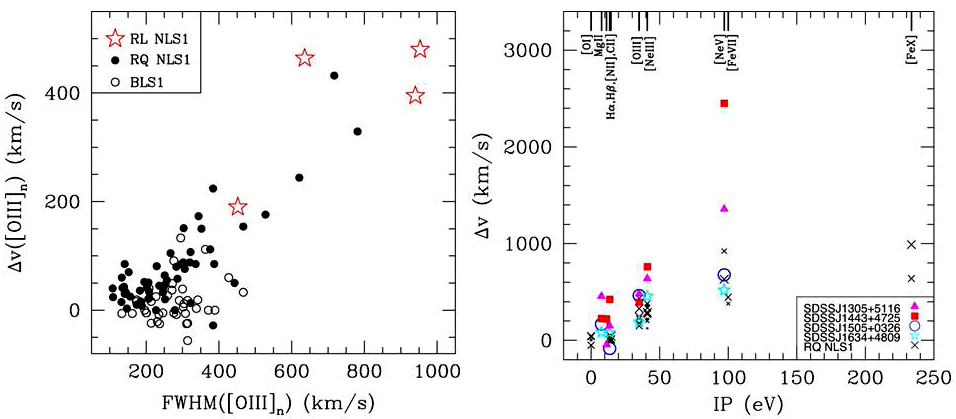}
\caption{Left: Strong correlation of [OIII] outflow velocity with line width, FWHM([OIII]), of the {\em core} component of [OIII] (blue wings are
additionally present, and are even more highly shifted; not plotted here). Radio-loud NLS1 galaxies are marked by red stars and show very high shifts and widths
(\cite{komossa18}). Right: Pronounced ionization stratification (radial velocity versus ionization potential IP). The symbols in colour represent
radio-loud NLS1 galaxies. For comparison, nine blue outliers among radio-quiet NLS1 galaxies (\cite{komossa08b}) are plotted with black crosses. }
\label{line-outflows}       
\end{figure*}

\subsection{Emission-line spectroscopy: High-velocity outflows}

The [OIII]$\lambda$5007 emission line is one of the most important spectral diagnostics in the optical band. It is important for classification of NLS1 galaxies, and acts as a valuable tracer of the physics and kinematics of the line-emitting gas. The [OIII] line profile can often be decomposed into two components: a core, and a fainter second Gaussian component called the ``wing''. These components come in two types: (1) Wings are often blueshifted and very common in AGN. However, (2) some sources have their whole [OIII] core line blueshifted. These have been termed ``blue outliers'' by \cite{zamanov02}.    
These are rare in BLS1 galaxies (at low $z$)\footnote{However, they are relatively abundant in powerful higher-redshift quasars; see \cite{marziani16} and references therein.}, but common in NLS1s. It has been found that 

\begin{itemize}
\item {About 16\% of all radio-quiet NLS1 galaxies are blue outliers and exhibit strong kinematic shifts of their [OIII] emission-line cores
($v > 150$ km s$^{-1}$, \cite{komossa08b}){\footnote{Because of the strong correlation of [OIII]$_{\rm core}$ line width with outflow velocity (Fig. \ref{line-outflows}), [OIII] width is no suitable surrogate for stellar velocity dispersion $\sigma_*$ in these systems (\cite{komossa07}), since it appears to be driven by outflows rather than the host bulge potential.}}. }

\item{The fraction of blue outliers in radio-loud NLS1 galaxies is slightly higher: $\sim$ 17-23\% (\cite{yuan08}, \cite{berton16b}, \cite{komossa18}), and the projected outflow velocities in the [OIII]-emitting gas are higher (Fig. \ref{line-outflows}).  

\item{$\gamma$-ray NLS1 galaxies show the highest fraction of [OIII] complexity (\cite{berton16b}).
}

\item{Very high outflow velocities have been detected in several radio-loud NLS1 galaxies in [NeV]$\lambda$3426, with projected velocities up to 2450 km/s (\cite{komossa18}), well above the escape velocities of the host galaxies. Three out of the four sources are candidate $\gamma$-ray emitters.}

}
\end{itemize}

While many studies of emission lines in NLS1 galaxies concentrated on [OIII] only, which is often one of the strongest features, additional
valuable diagnostics are provided by other emission lines including [OI]$\lambda$6300, [NeV]$\lambda$3426 and [FeVII]$\lambda$6087, which span a wide range in ionization potential and critical density.   
An analysis of all optical emission lines in four radio-loud NLS1 galaxies (\cite{komossa18}; our Tab. 3) -- one of them a $\gamma$-ray emitter (\cite{abdo09b}),
and two of them candidate $\gamma$-ray emitters (\cite{liao15}) -- has shown extreme line shifts which imply radial velocities exceeding 2000 km/s in the 
highest-ionization gas. Further, a strong ionization stratification is present along with a lack of zero-velocity high-ionization gas which implies that a large
NLR fraction is affected by the outflow. All galaxies of the mini-sample show high $L/L_{\rm Edd}$ and also
harbor strong radio jets. Therefore, both 
mechanisms -- large-scale winds and
jet-NLR interactions -- are potential drivers of the outflow.  
A few key properties of the sources are listed in
Tab. \ref{outflows}.  

\begin{table*}[t]
\small
\begin{center}
\begin{tabular}{lcccccccc}
\hline
galaxy & $z$ & log $R$ & $\Delta v_{\rm NeV}$ & $\Delta 
v_{\rm OIII_c}$ & $M_{\rm BH,H\beta}$ & $L/L_{\rm Edd}$ & log $P_{\rm jet}$ & $M_{\rm out}$ \\
       &     &         & km/s                  & km/s  &  10$^7$ M$_{\odot}$ &  & erg/s & 10$^7$ M$_{\odot}$ \\ 
\hline
\hline
SDSSJ1305+5116 & 0.785 & 2.34 & 1360 & 480 & 25 & 0.95 & 44.7 & 8.7 \\
SDSSJ1443+4725 & 0.703 & 3.07 & 2450 & 400 & 4.2 & 0.79 & 44.8 & 1.6 \\
PKS\,1502+0326  & 0.409 & 3.19 &  680 & 460 & 0.6 & 0.66 & 44.6 & 0.3 \\
\hline
\end{tabular}
\caption{Radio-loud NLS1 galaxies with the highest projected outflow velocities in [NeV]. $R$ is the radio-loudness index, $M_{\rm BH,H\beta}$ the H$\beta$-based SMBH mass, $P_{\rm jet}$ the jet power (estimated with equ. 16 of \cite{birzan08}), and $M_{\rm out}$ the ionized gas mass in outflow (based on [OIII], and with equ. 9 of \cite{komossa18}).    
\newline
}
\label{outflows}
\end{center}
\end{table*}

Based on predictions from hydrodynamical simulations of jets and winds expanding into the inhomogeneous interstellar medium of the host galaxy (\cite{wagner12}, \cite{wagner13}), and given the high jet powers of the sources, the high observed outflow velocities can be understood in a framework in which the radio-loud NLS1 galaxies are in an early stage of their evolution, not much older than 1 Myr (\cite{komossa18}). The pronounced ionization stratification across different ions and across a wide range of velocities remains exceptional and challenging to understand, but it, too, may arise in an early evolutionary stage. 
Spectroscopy of jetted NLS1 galaxies with extreme kinematic shifts of their optical emission lines, therefore, provides us with powerful tools for understanding high-velocity, large-scale outflows in Seyfert galaxies and quasars; their drivers, their association with radio jets, and their impact on the environment. 

\section{Summary}
In summary, there is growing evidence that radio-loud NLS1 galaxies are AGN in early stages of their evolution. They are actively evolving, are rapidly growing their black holes at high (near-Eddington) accretion rates, and are launching powerful jets which are still evolving across the inner regions of their host galaxies. They are, therefore, powerful laboratories {\em in the local} universe for understanding the physical processes which power
high-redshift quasars in the distant/early universe, {\em{and}} in particular for understanding the physics of blazar jets in a different parameter space and especially at shorter timescales than probed before.

\vskip0.4cm
{\itshape{Acknowledgments}}: It is my great pleasure to thank the organizers for this excellent meeting, and R. Antonucci, M. Berton, E. Bon, L. Foschini, L. Gallo, M. Gaskell, D. Grupe, P. Marziani, S. Mathur, H. Pan, B. Peterson, A. Wagner, M. Ward, D. Xu, H. Yang, S. Yao, W. Yuan, and H. Zhou for
enlightening discussions on NLS1 galaxies, and Nicholas MacDonald and an anonymous referee for their very useful comments on the manuscript. This conference has been organized with the support of the Department of Physics and Astronomy
``Galileo Galilei'', the University of Padova, the National Institute of Astrophysics INAF, the
Padova Planetarium, and the RadioNet consortium. RadioNet has received funding from the European
Union's Horizon 2020 research and innovation programme under grant agreement No~730562.

\end{document}